\documentclass{article}

\usepackage{arxiv}

\usepackage[utf8]{inputenc} 
\usepackage[T1]{fontenc}    
\usepackage{hyperref}       
\usepackage{url}            
\usepackage{booktabs}       
\usepackage{amsfonts}       
\usepackage{nicefrac}       
\usepackage{microtype}      
\usepackage{lipsum}
\usepackage{graphicx}

\usepackage{tikz, graphicx, mathtools, amsfonts, mathrsfs, amsmath, xfrac, amssymb, dsfont, enumitem,bbold,setspace, amsthm}

\usepackage[ruled,vlined]{algorithm2e}
\SetKwInOut{Parameters}{Parameters}
\SetKwInOut{Initialization}{Initialization}
\SetAlgoSkip{bigskip}

\newcommand{\reals}{\mathbb{R}}

\newcommand{\diam}[1]{\mathrm{diam}\left(#1\right)}
\newcommand{\diag}[1]{\mathrm{diag}\left\{#1\right\}}
\newcommand{\abs}[1]{\left\lvert#1\right\rvert}
\newcommand{\adjm}[1]{\mathrm{Adj}\left[#1\right]}
\newcommand{\degm}[1]{\mathrm{D}\left[#1\right]}
\newcommand{\norm}[1]{\left\lVert#1\right\rVert}
\newcommand{\proj}[2]{\mathrm{Proj}_{#1}\left[#2\right]}

\usepackage{listings,bbold}
\usepackage{color}

\DeclareMathOperator*{\argmax}{arg\,max}

\renewcommand{\abs}[1]{\left\lvert#1\right\rvert}
\renewcommand{\norm}[2]{\left\lVert#1\right\rVert_{#2}}

\newtheorem{remark}{Remark}
\newtheorem{lemma}{Lemma}
\newtheorem{theorem}{Theorem}
\newtheorem{assume}{Assumption}
\newtheorem{define}{Definition}

\title{An Algorithm for Distributed Computation of Reachable Sets for Multi-Agent Systems
\thanks{This work has been submitted to American Control Conference 2025 for possible publication.}}

\author{
 Omanshu Thapliyal \\
  UAM Researcher\\
  Big Data Analytics \& Solutions Lab\\
  Hitachi America Ltd. \\
  \texttt{omanshu.thapliyal@hal.hitachi.com} \\
   \And
 Shanelle Clarke \\
  Sr. Autonomy Research Engineer\\
  Supernal\\
  \texttt{Shanelle.Clarke@supernal.aero} \\
  \And
 Inseok Hwang \\
  Professor, Aeronautics and Astronautics\\
  School of Aeronautics and Astronautics\\
  Purdue University\\
  \texttt{ihwang@purdue.edu} \\
}

\begin{document}

\maketitle

\begin{abstract}
In this paper, we consider the problem of distributed reachable set computation for multi-agent systems (MASs) interacting over an undirected, stationary graph.
A full state-feedback control input for such MASs depends no only on the current agent's state, but also of its neighbors.
However, in most MAS applications, the dynamics are obscured by individual agents.
This makes reachable set computation, in a fully distributed manner, a challenging problem.
We utilize the ideas of polytopic reachable set approximation and generalize it to a MAS setup.
We formulate the resulting sub-problems in a fully distributed manner and provide convergence guarantees for the associated computations.
The proposed algorithm's convergence is proved for two cases: static MAS graphs, and time-varying graphs under certain restrictions.
\end{abstract}

\section{Introduction}
\label{sec:introduction}
Multi-agent systems (MASs) appear in widely different fields, ranging from social networks, biological systems, computer networks, smart grid, to multi-robot systems \cite{chen2019control}.
Such systems often consist of cheaper or smaller components with limited capabilities, to perform complex tasks that an individual agent could not perform otherwise -- such as formation control, target tracking and capture, distributed computing, and distributed learning tasks \cite{dorri2018multi}.
Multi-robot systems are employed to enhance capabilities of individual units by virtue of the additional redundancy, resilience, robustness against agent failure, and modular task assignment.
As a result, the emergent behavior of multi-agent systems can result in behaviors far more complex than the comprising individual agents.
Due to the above benefits of MASs, such systems find ubiquitous applications, often in safety-critical environments \cite{iocchi2000reactivity}.
Therefore, the ability to compute safety properties on-line for MASs is highly desirable.

\begin{figure}[!ht]
    \centering
    \resizebox{1\columnwidth}{!}
    {
    \input{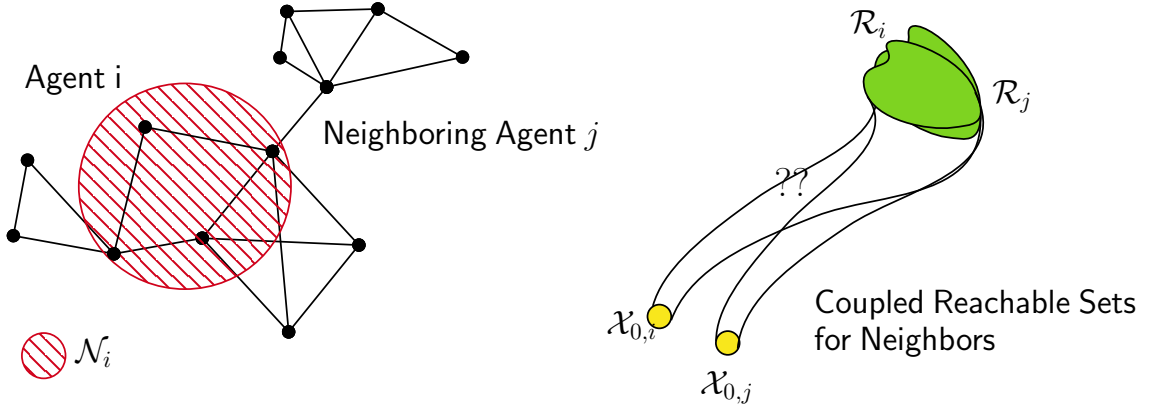}
    }
    \caption{Dynamical coupling of reachable sets in an MAS}
    \vspace{-0.0em}
    \label{fig:distr-reach}
\end{figure}

Of the various set-properties that characterize safety, the ability to compute reachability properties for dynamical systems is an important task for control synthesis, validation of control protocols, and online safety checking.
For the more elaborate dynamical systems with dedicated computational resources, a centralized reachability problem is limited only by the accuracy of the dynamical model.
However, MAS dynamics have to be flexible enough to allow for transient collaboration, followed by periods of non-cooperative, `selfish' dynamics \cite{oh2015survey,ren2005survey,horling2004survey}.
Additionally, MASs are usually cheaper components in a bigger network of agents, internet-of-things network, or a system-of-systems, with limited computational capabilities.
The limited computational capability at each agent is the most severe bottleneck in computing properties of an MAS in a distributed manner.
Further, the sensed and communicated information from within the neighborhood of any agent can affect its own reachable sets (therefore, its own safety property) in non-trivial ways (see Fig.~\ref{fig:distr-reach}).
The distributed reachable set computation problem is further exacerbated by the individual vehicular dynamics being obscured across the MAS network of interacting UAVs.

Reachability problems for more elaborate dynamical systems have been dealt with in literature in a centralized manner \cite{hwang2005polytopic,varaiya2000reach,thapliyal2022approximate,bansal2016learning,bansal2017hamilton,dit2018reachability,kwon2017reachability,thapliyal2023approximating, faure2022survey,kurzhanski2000ellipsoidal, thapliyal2023data}.
Computing exact reachable sets often requires solving Hamilton Jacobi (HJ) equations \cite{bansal2016learning,bansal2017hamilton}, or finding level set functions \cite{hwang2005polytopic} to represent set boundaries.
On the other hand, authors in \cite{kwon2017reachability,varaiya2000reach,thapliyal2022approximate} provide centralized reachable set approximation methods by bounding the exact reachable sets with ellipses, polytopes, and intervals, respectively.
However, a centralized reachability framework does not take into account limited computational capabilities at each agent.
Neither does a centralized reachability framework address the limited information sharing structure of decentralized MASs.
A few works in the literature consider multi-agent applications of reachability \cite{kariotoglou2015multi,wang2020infusing,wang2021formal,share2017step,tian2021reachable}.
Authors in \cite{kariotoglou2015multi} utilize centralized definitions of reachability applied to certain cases of multi-agent networks.
The actual computation of the reachable sets does not require communication of information, or is dependent on the states of neighboring agents.
In \cite{wang2021formal} a multi-agent reinforcement learning (RL) framework is considered where the overall safety requirements are written in a formal mixed-integer linear programming (MILP) notation.
However, a central MILP is solved to compute reachable sets for the multi-agent RL framework.
Methods proposed in \cite{wang2020infusing} utilize Hamilton Jacobi reachability, therefore, suffer from the well known `curse of dimensionality'.
Authors in \cite{share2017step, tian2021reachable} consider similar centralized variations of the reachable set computation problem.
However, none of the existing methods address the limited computational abilities, or the couple dynamical models of individual agents.
Further, the agents cooperate in a mission inflexible manner, and little allowance is given to graph connectivity (except \cite{tian2021reachable}).

To this end, we attempt to extend the ideas of approximate reachable set computation, and implement a fully distributed algorithm that incorporates the inter-agent dynamical coupling behavior.
The proposed algorithm has convergence guarantees for static inter-agent networks that are strongly connected.
Our main contributions are: (i) posing reachable set computation for MASs as a distributed problem, (ii) proposing a fully distributed algorithm to compute tight approximations to the reachable sets, and (iii) providing convergence guarantees for the proposed algorithm for both cases: static and  time-varying graphs and discuss in detail the requirements for convergence.
Furthermore, the resulting algorithm is computationally inexpensive, does not require solving MILPs or HJ equations, and the required computations can be easily computed locally by individual agents.


The remainder of this paper is organized as follows.
In Section \ref{sec:problemstatement} we detail the distributed reachability problem.
Section \ref{sec:methodology} contains the proposed methodology to solve the reachability problem for multi-agent systems in a fully distributed manner.
Section \ref{sec:convergence} presents a discussion on the convergence of the complete algorithm for distributed reachable set computation in detail.
We consider both cases for the MAS connectivity graphs: static graphs, and time-varying graphs, and outline the graph requirements for convergence.
Finally, in Section \ref{sec:conclusion} we present our concluding remarks and future directions.

\section{Problem Formulation}\label{sec:problemstatement}
Consider a multi-agent system (MAS) where individual agent states are coupled by a state feedback control law, with multiplicative gains for each network in the agents as:
\begin{equation}\label{eq:system-eqn}
\begin{split}
\dot{x}_i(t) &= A_i x_i(t) + B_i u_i(t) + B_{1,i} w_i(t)
\end{split}
\end{equation}
\begin{equation}\label{eq:coupled-control}
\text{where } u_i(t) = K_{ii}x_i(t) + \sum_{j\in\mathcal N_i}{K_{ij} (x_i(t)-x_j(t))}    
\end{equation}
Here the state of agent $i$ is $x_i\in\reals^{n_x}$, the control input $u_i\in\reals^{n_u}$, and an unknown, exogenous control $w_i\in\reals^{n_w}$, with gain matrices of appropriate sizes $K_i$, and system matrices $A_i,B_i$ and $B_{1,i}$.
The agents can communicate over a network denoted by the undirected graph $\mathcal G = (\mathcal V,\mathcal E)$, where $\mathcal V=\{1,\dots,N\}$ is the set of agents, and $\mathcal E\subseteq\{(i,j):i,j\in\mathcal V\}$ is the set of all edges.
The problem is characterized by agents having their dynamics coupled directly through their neighboring agents' states $x_j\in\mathcal N_i$ feeding into their own control inputs to modify state $x_i$ from (\ref{eq:system-eqn}).
Additionally, the dynamics of different agents are unknown to each other, i.e., $(A_i,B_i)$ is known only to the agent $i$ themselves.

Given the dynamical system in (\ref{eq:system-eqn}), we can now define the reachable set computation problem.
The reachable set for agent $i\in\mathcal V$ is defined as the set of all possible states for agent $i$ at some time $\tau\geq t_0$ as:
\begin{equation}\label{eq:reach-i}
\begin{split}
\mathcal R_i(\tau;\mathcal X_{0,i},\mathcal W_i) \triangleq \{&x(\tau)\in\reals^{n_x}:\dot{x}=A_i x+B_iu_i+B_{1,i}w_i,\\
& w_i(t)\in\mathcal W_i,x(t_0)\in\mathcal X_{0,i}\}
\end{split}
\end{equation}
for initial set $\mathcal X_{0,i}$, and admissible exogenous control set $\mathcal W_i$.
\begin{assume}
Let the exogenous input set be bounded as $\mathcal W_i\triangleq \{w_i(t)\in\reals^{n_w}\,\lvert\,\norm{w_i(t)}{2}\leq \rho_i\}$. 
\end{assume}

In case the assumption above does not hold, the set $\mathcal W$ can easily be bounded by a sphere of radius $\rho_i$ for a finite set $\mathcal W$.
From the `neighborhood-state feedback' control from (\ref{eq:coupled-control}), the reachable set  $\mathcal R_i(\tau;\mathcal X_{0,i},\mathcal W_i)$ clearly depends on $x_j\,\forall j\in\mathcal N_i$.
Simultaneously, $x_j$ also evolves under dynamics $(A_j,B_j)$.
The distributed reachability problem is to compute the reachable sets in (\ref{eq:reach-i}) for each agent, under coupled control input in (\ref{eq:coupled-control}), such that the dynamics $(A_i,B_i)$ are not shared among agents.

\section{Distributed Reachable Set Computation}\label{sec:methodology}
To deal with the distributed reachability problem above, let us consider a simpler scenario.
Let $\mathcal L$ be the Laplacian of the graph corresponding to $\mathcal G$, i.e.,
\begin{equation}\label{eq:laplacian}
\mathcal L_{\mathcal G} \triangleq \degm{\mathcal G} - \adjm{\mathcal G}
\end{equation}
where $\degm{\bullet}$ and $\adjm{\bullet}$ are the graph degree and adjacency matrices, respectively.
Before we attempt to solve the distributed reachability problem for the MAS, let us consider a centralized approach.
To this end, we define the dynamics of the \textit{stacked system} for the centralized representation of the MAS by redefining the following variables: 
\begin{equation}\label{eq:stacked-var-def}
\begin{split}
\xi(t) &\triangleq [x_1(t)^T,\cdots,x_N(t)^T]^T,\\
W(t) &\triangleq [w_1(t)^T,\cdots,w_N(t)^T]^T\\
\mathbb{A}(t) &\triangleq\diag{A_i+B_iK_{ii}}_{i\in\mathcal V} + \mathcal L_\mathcal{G} \otimes B_i K_{ij} , \text{ and}\\
\mathbb B &\triangleq \diag{B_{1,i}}_{i\in\mathcal V}
\end{split}    
\end{equation}
This allows us to consider the evolution of the centralized MAS states $\xi(t)\in\reals^{n_x\times N}$ as:
\begin{equation}\label{eq:concatenated-evolution}
\dot{\xi}(t) = \mathbb A(t) \xi(t) + \mathbb B W(t)
\end{equation}
The stacked system in (\ref{eq:stacked-var-def}),~(\ref{eq:concatenated-evolution}) allows us to consider a simpler, centralized reachability problem.

\subsection{Centralized Reachable Set Computation}
If a central computing entity were to compute the reachable sets $\mathcal R_i(\tau;\mathcal X_{o,i},\mathcal W_i)$, the equivalent centralized reachability problem can written compactly as:
\begin{subequations}\label{eq:central-reach-prob}
\begin{align}
\mathcal R^\xi(\tau;\Xi_0, \mathcal W) &\triangleq \Big\{\xi(\tau) : \dot{\xi}(t) = \mathbb A \xi(t) + \mathbb B W(t), \tau\geq t_0 ,\nonumber \\
&\qquad \qquad \qquad \xi_i(t_0)\in \Xi_0 , W(t)\in\mathcal W \Big\} \\
\text{where }\Xi_0 &\triangleq \prod_{i=1}^N \mathcal X_{i,0} = \mathcal X_{1,0}\times\cdots\times \mathcal X_{N,0}\,, \text{ and }\nonumber \\
\mathcal W &\triangleq \prod_{i=1}^N \mathcal W_i = \mathcal W_{1}\times\cdots\times \mathcal W_{N}
\end{align}
\end{subequations}

From (\ref{eq:central-reach-prob}), the centralized reachability problem for the MAS is clearly a linear-system reachable set computation problem.
Note that if the graph $\mathcal G$ is static (e.g., a smart grid, internet network, robot flocking, etc.), centralized system matrix $\mathbb A(t) = \mathbb A$ is time invariant.
For simplicity of analysis, we will initially concern ourselves with the stationary graph such that $\mathcal G(t) = \mathcal G$.
Therefore, we first investigate reachable set computation for the linear time invariant (LTI) system $(\mathbb A, \mathbb B)$ given characterizations of initial central state set $\Xi_0$ and admissible exogenous input set $\mathcal W$.

\begin{figure*}[!htb]
    \centering
    \resizebox{0.975\textwidth}{!}
    {
  
\tikzset {_6tgfyomj9/.code = {\pgfsetadditionalshadetransform{ \pgftransformshift{\pgfpoint{89.1 bp } { -108.9 bp }  }  \pgftransformscale{1.32 }  }}}
\pgfdeclareradialshading{_8z8rqnw52}{\pgfpoint{-72bp}{88bp}}{rgb(0bp)=(1,1,1);
rgb(0bp)=(1,1,1);
rgb(25bp)=(0,0,0);
rgb(400bp)=(0,0,0)}

  
\tikzset {_jfzmwe045/.code = {\pgfsetadditionalshadetransform{ \pgftransformshift{\pgfpoint{89.1 bp } { -128.7 bp }  }  \pgftransformscale{1.32 }  }}}
\pgfdeclareradialshading{_2cswvzbq1}{\pgfpoint{-72bp}{104bp}}{rgb(0bp)=(1,1,1);
rgb(0bp)=(1,1,1);
rgb(19.196428571428573bp)=(0.48,0.15,0.15);
rgb(400bp)=(0.48,0.15,0.15)}

  
\tikzset {_2vh0k6jjw/.code = {\pgfsetadditionalshadetransform{ \pgftransformshift{\pgfpoint{89.1 bp } { -108.9 bp }  }  \pgftransformscale{1.32 }  }}}
\pgfdeclareradialshading{_h6p5guust}{\pgfpoint{-72bp}{88bp}}{rgb(0bp)=(1,1,1);
rgb(0bp)=(1,1,1);
rgb(25bp)=(0,0,0);
rgb(400bp)=(0,0,0)}
\tikzset{every picture/.style={line width=0.75pt}} 

\begin{tikzpicture}[x=0.75pt,y=0.75pt,yscale=-1,xscale=1]

\draw  [draw opacity=0][shading=_8z8rqnw52,_6tgfyomj9] (85.33,267.6) .. controls (85.33,249.37) and (106.6,234.6) .. (132.83,234.6) .. controls (159.07,234.6) and (180.33,249.37) .. (180.33,267.6) .. controls (180.33,285.83) and (159.07,300.6) .. (132.83,300.6) .. controls (106.6,300.6) and (85.33,285.83) .. (85.33,267.6) -- cycle ;
\draw  [draw opacity=0][shading=_2cswvzbq1,_jfzmwe045] (242.67,140.27) .. controls (242.67,113.94) and (273.48,92.6) .. (311.5,92.6) .. controls (349.52,92.6) and (380.33,113.94) .. (380.33,140.27) .. controls (380.33,166.59) and (349.52,187.93) .. (311.5,187.93) .. controls (273.48,187.93) and (242.67,166.59) .. (242.67,140.27) -- cycle ;
\draw  [fill={rgb, 255:red, 155; green, 155; blue, 155 }  ,fill opacity=0.39 ][dash pattern={on 4.5pt off 4.5pt}] (126.71,182.38) -- (175.37,214.35) -- (71.98,326.3) -- (23.32,294.33) -- cycle ;
\draw    (97.67,257.27) -- (132.83,267.6) ;
\draw [shift={(97.67,257.27)}, rotate = 16.37] [color={rgb, 255:red, 0; green, 0; blue, 0 }  ][fill={rgb, 255:red, 0; green, 0; blue, 0 }  ][line width=0.75]      (0, 0) circle [x radius= 3.35, y radius= 3.35]   ;
\draw  [fill={rgb, 255:red, 248; green, 231; blue, 28 }  ,fill opacity=0.2 ][dash pattern={on 4.5pt off 4.5pt}] (236.66,49.32) -- (306.09,65.74) -- (245.42,189.36) -- (176,172.93) -- cycle ;
\draw    (257,123.93) -- (311.5,140.27) ;
\draw [shift={(257,123.93)}, rotate = 16.68] [color={rgb, 255:red, 0; green, 0; blue, 0 }  ][fill={rgb, 255:red, 0; green, 0; blue, 0 }  ][line width=0.75]      (0, 0) circle [x radius= 3.35, y radius= 3.35]   ;
\draw [line width=1.5]  [dash pattern={on 5.63pt off 4.5pt}]  (99.35,254.34) .. controls (213,231.93) and (213.67,153.27) .. (253.67,123.27) ;
\draw [line width=1.5]  [dash pattern={on 1.69pt off 2.76pt}]  (132.83,267.6) .. controls (311.67,242.6) and (226.33,198.6) .. (307.67,140.6) ;
\draw  [draw opacity=0][shading=_h6p5guust,_2vh0k6jjw] (477.7,221.25) .. controls (471.45,198.41) and (495.14,172.03) .. (530.62,162.32) .. controls (566.09,152.61) and (599.92,163.25) .. (606.17,186.08) .. controls (612.42,208.91) and (588.73,235.29) .. (553.25,245) .. controls (517.77,254.71) and (483.95,244.08) .. (477.7,221.25) -- cycle ;
\draw   (557.39,173.78) -- (501.83,202.59) -- (478.67,191.58) -- (519.92,155.98) -- (568.57,144.98) -- cycle ;
\draw   (501.83,202.59) -- (557.4,173.78) -- (562.95,215.61) -- (534.47,249.08) -- (491.8,247.07) -- cycle ;
\draw   (534.47,249.08) -- (562.95,215.61) -- (579.37,206.61) -- (575,246.07) -- (552.6,264.17) -- cycle ;
\draw   (475.8,229.98) -- (475.8,229.98) -- (478.66,191.58) -- (501.83,202.59) -- (491.8,247.07) -- cycle ;
\draw   (520.6,270.2) -- (491.8,247.07) -- (534.47,249.08) -- (534.47,249.08) -- (552.6,264.17) -- cycle ;
\draw   (593.13,167.63) -- (568.58,144.98) -- (593.13,167.63) -- (593.13,167.63) -- (579.37,206.61) -- cycle ;
\draw   (619.8,192.77) -- (593.13,167.63) -- (579.37,206.61) -- (575,246.07) -- (619.8,192.77) -- cycle ;

\draw  [line width=2.25]  (446.4,95.8) -- (634.2,95.8) -- (634.2,341.4) -- (446.4,341.4) -- cycle ;

\draw (132.83,300.6) node [anchor=north west][inner sep=0.75pt]  [font=\large]   {$\Xi _{0}$};
\draw (360.67,14.93) node [anchor=north west][inner sep=0.75pt]  [font=\large]   {$\mathcal{R}^{\xi }( \tau ;\Xi _{0})$};
\draw (66,243.6) node [anchor=north west][inner sep=0.75pt]  [font=\large]   {$\xi _{0j}^{*}$};
\draw (48.67,166.93) node [anchor=north west][inner sep=0.75pt]  [font=\large]   {$\left( c_{oj} ,\gamma _{0j}^{*}\right)$};
\draw (30,129.6) node [anchor=north west][inner sep=0.75pt]  [font=\large]   {$j^{th} \ \mathsf{Hyperplane}$};
\draw (94.67,327.33) node [anchor=north west][inner sep=0.75pt]  [font=\large]   {$\mathsf{Initial\ Set}$};
\draw (360.67,61.6) node [anchor=north west][inner sep=0.75pt]  [font=\large]   {$ \begin{array}{l}
\mathsf{Reachable\ Set}\\
\mathsf{at\ the\ \tau }
\end{array}$};
\draw (204.67,106.27) node [anchor=north west][inner sep=0.75pt]  [font=\large]   {$\xi _{j}^{*}( t)$};
\draw (244.33,240.93) node [anchor=north west][inner sep=0.75pt]  [font=\large]   {$\dot{\xi }( t) =\mathbb{A} \xi ( t) \ +\ W( t)$};
\draw (473.87,267.33) node [anchor=north west][inner sep=0.75pt]  [font=\large]   {$\bigcap _{j=1}^{n_{h}}\left\{\xi \in \left( c_{0j} ,\gamma _{0j}^{*}\right)\right\}$};
\draw (452.27,101.73) node [anchor=north west][inner sep=0.75pt]  [font=\large]   {$\mathsf{Polytopic\ Approximation}$};

\end{tikzpicture}
    }
    \caption{Polytopic Approximation of Reachable Set: Evolution of a selected Hyperplane}
    \vspace{-0.0em}
    \label{fig:plane-evolve}
\end{figure*}
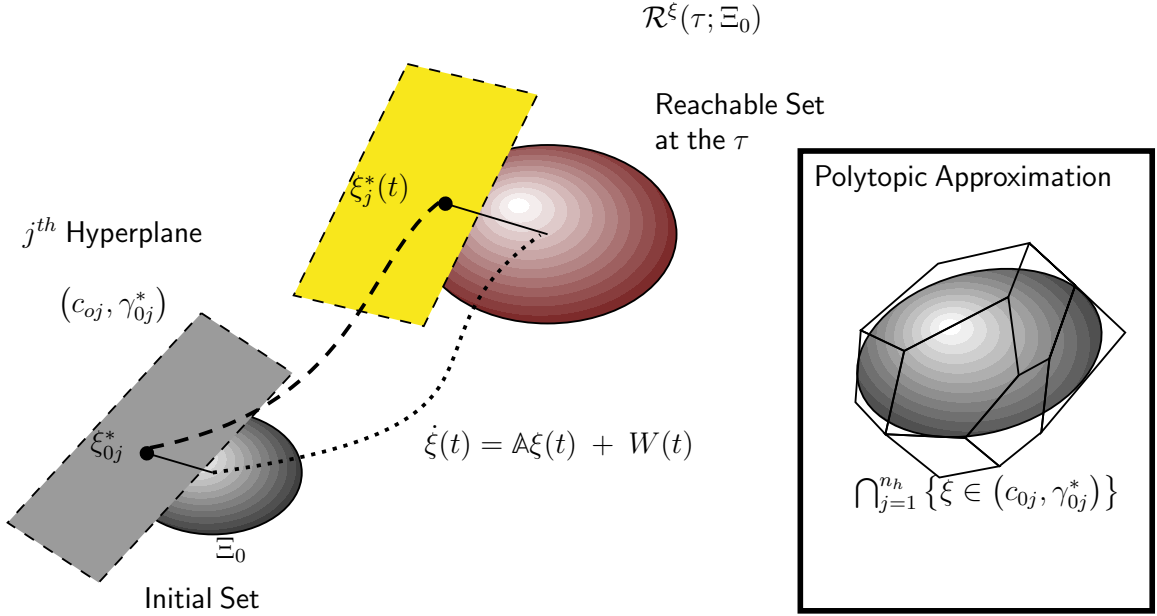

\begin{lemma}
The admissible exogenous input set $\mathcal W$ is convex and bounded.
\end{lemma}
\begin{proof}
Let $v_1,v_2\in\mathcal W$.
Let $v_1=[a_1,\cdots,a_n]^T$ and $v_2=[b_1,\cdots,b_n]^T$.
From Assumption 1, we have $\norm{a_i}{2}\leq \rho_i$ and $\norm{b_i}{2}\leq \rho_i$.
Consider their convex combination 
\begin{equation*}\label{eq:convex1}
\alpha v_1 + (1-\alpha) v_2 = \begin{bmatrix}
\alpha a_1 + (1-\alpha) b_1 \\ \vdots \\ \alpha a_n + (1-\alpha) b_n
\end{bmatrix}
\end{equation*}
for some $\alpha\in(0,1)$.
Then, the following holds: $\norm{\alpha a_j + (1-\alpha) b_j}{2} \leq \alpha \norm{a_j}{2} + (1-\alpha)\norm{b_j}{2} \leq \alpha \rho_j + (1-\alpha) \rho_j = \rho_j$.
So the convex combination of $v_1,v_2$ also lies in $\mathcal W$, also, the set $\mathcal W$ is bounded.
\end{proof}
\vspace{-0.0em}
We will concern ourselves with polytopic approximations of $\Xi_0$ and $\mathcal W$ utilizing approximate reachable set computation from \cite{varaiya2000reach, thapliyal2023approximating}.
Without a loss of generality, consider the initial state set and the admissible exogenous input set as:

\begin{align}
\Xi_0 &= \bigcap_{j=1}^{n_{1}} {\{v\in\reals^{Nn_x}\,: \,\langle c_j(t_0), v\rangle \leq \gamma_j(t_0)\}} \label{eq:polytope-xi-def} \\ 
\mathcal W &= \bigcap_{j=1}^{n_{2}} {\{u\in\reals^{Nn_u}\, : \,\langle d_j, u\rangle \leq \varepsilon_j\}}  \label{eq:polytope-w-def}
\end{align}
where the polytopic sets $\Xi_0$ and $\mathcal W$ have $n_1$ and $n_2$ faces, respectively, and variables $c_j(t_0),\gamma_j(t_0),d_j,\varepsilon_j$ parameterize the hyperplanes defining the faces of the polytopes.
Thus, the $j^\text{th}$ hyperplane supporting the convex set $\Xi_0$ can be denoted by the tuple $H_j\triangleq (c_j(t_0),\gamma_j(t_0))$, as shown in Fig.~\ref{fig:plane-evolve}.

\begin{theorem}[Polytopic Reachability \cite{varaiya2000reach}]\label{th:poly-reach1}
Let hyperplane $H_j=(c_j(t_0),\gamma_j(t_0))$ support the initial set $\Xi_0$ at time $t_0$, at a point $\xi_j^*(t_0)$.
Then the point of contact $\xi_j^*(t_0)$ evolves as:
\begin{align}\label{eq:support-point-evolve}
\dot{\xi}_j(\tau) &= \mathbb A \xi_j(\tau) + \mathbb B W_j^*(\tau)
\end{align}
where $W_j^*(\tau)$ solves the optimal control problem:
\begin{align}
W_j^*(\tau) &= \underset{W\in\mathcal W}{\argmax} {\left\{ \langle \lambda_j^*(\tau), \mathbb A \xi_j^*(\tau) + \mathbb B W\rangle \right\}} \label{eq:optimal-control} \\    
\dot{\lambda}_j^*(\tau) &= -\mathbb A^T{\lambda}_j^*(\tau)\text{ s.t. }, {\lambda}_j^*(t_0) = c_j(t_0), \,t_0\leq\tau\leq t \label{eq:costate-evolve}
\end{align}
for the $j^\text{th}$ hyperplane, and the co-state variable $\lambda_j^*(\tau)$.
Additionally, for $\gamma^*_j(\tau) \triangleq \langle \lambda^*_j(\tau), \xi_j^*(\tau) \rangle$, the hyperplane $(\lambda_j^*(\tau),\gamma^*_j(\tau))$ supports the reachable set $\mathcal R^\xi(\tau;\Xi_0,\mathcal W)$ at time $\tau$, at the point $\xi^*_j(\tau)$.
\end{theorem}
\begin{proof}
Proof follows from Varaiya et al. in \cite{varaiya2000reach} and Theorem 1 in \cite{thapliyal2023approximating}. 
\end{proof}

A geometric representation of Theorem 1 is depicted in Fig.~\ref{fig:plane-evolve}, showing the evolution of the point of contact of hyperplane $H_j$ over time.
This allows us to find the evolution of the supporting points of all $n_1$ hyperplanes by solving equations (\ref{eq:support-point-evolve}) through (\ref{eq:costate-evolve}) for each $j$ in $1\leq j\leq n_1$.
As a direct consequence of Theorem 1, the reachable set is bounded by the polytope:
\begin{equation}\label{eq:reach-bound-polytope}
\mathcal R^\xi(\tau) \subset \bigcap_{j=1}^{n_1} {\left\{ \xi : \langle\lambda_j^*(\tau) ,\xi\rangle \leq \gamma _{j}^{*}(\tau)\right\}}
\end{equation}
where the hyperplane $(\lambda_j^*(\tau),\gamma _{j}^{*}(\tau))$ touches the reachable set at the point $\xi^*_j(\tau)$, as shown in Fig.~\ref{fig:plane-evolve}.

Suppose $\dot{z}=M(t)z + B(t)u(t)$ is an arbitrary linear dynamical system.
We define the state transition matrix the (at some time $t$) linear system $(M,B)$ as $\Phi_{ M}(t,t_0)$ such that $\dot{\Phi_{M}(t,t_0) = M(t)\Phi_{M}(t,t_0)}$, and $\Phi_{M}(t_1,t_1)=I$.
Therefore, due to Theorem \ref{th:poly-reach1}, the centralized computation of polytopic approximations of the reachable set can be carried out using the following steps.
\begin{enumerate}[leftmargin=*, label=S\arabic*:]
\item Initialize $\lambda^*_j(t_0) = c_{j}(t_0)$, then propagate the co-state as $\lambda^*_j(\tau) = \Phi_{-\mathbb{A}^T}(\tau,t_0) \lambda^*_j(t_0)$ from (\ref{eq:costate-evolve}).
\item Compute $W_j^*(\tau)$ in (\ref{eq:optimal-control}) using the co-state from step 1:\\
$ W_j^*(\tau) = \underset{W\in\mathcal W}{\argmax}{\left\{ \langle \Phi_{-\mathbb{A}^T}(\tau,t_0) \lambda^*_j(t_0), W\rangle \right\}} $.
\item Finally, the contact point is propagated using (\ref{eq:support-point-evolve}) as:
\begin{equation}
\xi^*_j(\tau) = \Phi_{\mathbb{A}}(\tau,t_0)\xi_{j}(t_0)^* + \int^\tau_{t_0} { \Phi_{\mathbb{A}}(\tau,s)\mathbb B W_j^*(s)} ds
\end{equation}
\end{enumerate}

It must be noted from \cite{varaiya2000reach}, that the polytopic representation in (\ref{eq:reach-bound-polytope}) simultaneously provides the outer and inner approximations to the reachable set as:
\begin{equation}\label{eq:outer-inner-approx}
\begin{split}
\text{convex hull} &\{\xi^*_i(\tau),\cdots,\xi^*_{n_1}(\tau)\} \subset \mathcal R^\xi(\tau),\text{ and} \\
\mathcal R^\xi(\tau) &\subset \bigcap_{j=1}^{n_1} {\left\{ \xi : \langle\lambda_j^*(\tau) ,\xi\rangle \leq \gamma _{j}^{*}(\tau)\right\}}
\end{split}
\end{equation}

\begin{remark}[Tightness of Approximation in (\ref{eq:outer-inner-approx})]
The approximation in (\ref{eq:outer-inner-approx}) is tight in the sense that there does not exist a polytope with $n_1$ verticeshttps://explorabl.es/all/ $\underline{P}_{n_1}$, or a polytope $\overline{P}_{n_1}$ with $n_1$ faces, such that $\text{convex hull} \{\xi^*_i(\tau),\cdots,\xi^*_{n_1}(\tau)\} \subset \underline{P}_{n_1} \subset \mathcal R^\xi(\tau) \subset \overline{P}_{n_1} \subset \bigcap_{j=1}^{n_1} {\left\{ \xi : \langle\lambda_j^*(\tau) ,\xi\rangle \leq \gamma _{j}^{*}(\tau)\right\}}$ (see \cite{varaiya2000reach}).
\end{remark}

\subsection{Distributed Reachable Set Algorithm}
Now that we established the procedure to compute centralized approximations (both inner, and outer) to the reachable set $\mathcal R^\xi(\tau)$ in steps S1-S3, we investigate the decentralized problem for the MAS in (\ref{eq:system-eqn}).
Therefore, the information available to agent $i$ can be summarized as:
\begin{equation}\label{eq:information}
\mathcal I_i \triangleq \big\{A_i,B_i,B_{1,i},\mathcal N_i, \{K_{ij}\}_{j\in\mathcal N_i},\rho_i, \{F_i^k\}_{k=1}^{n_1}, \{H_i^k\}_{k=1}^{n_2}\big\}    
\end{equation}
where $\{F_i^j\}_{j=1}^{n_1}$ and $\{H_i^j\}_{j=1}^{n_2}$ are the hyperplanes defining the bounding polytopes for local sets $\mathcal X_{0,i}$ and $\mathcal W_i$, respectively.
For simplicity of analysis, we assume the polytopes for each agent have the same number of faces $n_1$ and $n_2$.

\begin{figure}[!ht]
    \centering
    \resizebox{0.8\columnwidth}{!}
    {
    \tikzset{every picture/.style={line width=0.75pt}} 

\begin{tikzpicture}[x=0.75pt,y=0.75pt,yscale=-1,xscale=1]

\draw    (51,70) -- (194,70) ;
\draw [shift={(196,70)}, rotate = 180] [fill={rgb, 255:red, 0; green, 0; blue, 0 }  ][line width=0.08]  [draw opacity=0] (12,-3) -- (0,0) -- (12,3) -- cycle    ;
\draw [shift={(49,70)}, rotate = 0] [fill={rgb, 255:red, 0; green, 0; blue, 0 }  ][line width=0.08]  [draw opacity=0] (12,-3) -- (0,0) -- (12,3) -- cycle    ;
\draw    (85,75) -- (85,64) ;
\draw    (156,76) -- (156,65) ;
\draw    (50,139) -- (193,139) ;
\draw [shift={(195,139)}, rotate = 180] [fill={rgb, 255:red, 0; green, 0; blue, 0 }  ][line width=0.08]  [draw opacity=0] (12,-3) -- (0,0) -- (12,3) -- cycle    ;
\draw [shift={(48,139)}, rotate = 0] [fill={rgb, 255:red, 0; green, 0; blue, 0 }  ][line width=0.08]  [draw opacity=0] (12,-3) -- (0,0) -- (12,3) -- cycle    ;
\draw    (84,144) -- (84,133) ;
\draw    (155,145) -- (155,134) ;
\draw    (51,213) -- (194,213) ;
\draw [shift={(196,213)}, rotate = 180] [fill={rgb, 255:red, 0; green, 0; blue, 0 }  ][line width=0.08]  [draw opacity=0] (12,-3) -- (0,0) -- (12,3) -- cycle    ;
\draw [shift={(49,213)}, rotate = 0] [fill={rgb, 255:red, 0; green, 0; blue, 0 }  ][line width=0.08]  [draw opacity=0] (12,-3) -- (0,0) -- (12,3) -- cycle    ;
\draw    (85,218) -- (85,207) ;
\draw    (156,219) -- (156,208) ;
\draw  [color={rgb, 255:red, 74; green, 74; blue, 74 }  ,draw opacity=1 ][line width=2.25]  (203,79.25) -- (399.63,79.25) -- (399.63,44) -- (445,130) -- (399.63,216) -- (399.63,180.75) -- (203,180.75) -- cycle ;
\draw    (504,175) -- (643,182.89) ;
\draw [shift={(645,183)}, rotate = 183.25] [fill={rgb, 255:red, 0; green, 0; blue, 0 }  ][line width=0.08]  [draw opacity=0] (12,-3) -- (0,0) -- (12,3) -- cycle    ;
\draw    (504,175) -- (421.36,264.53) ;
\draw [shift={(420,266)}, rotate = 312.71] [fill={rgb, 255:red, 0; green, 0; blue, 0 }  ][line width=0.08]  [draw opacity=0] (12,-3) -- (0,0) -- (12,3) -- cycle    ;
\draw    (504,175) -- (503.01,36) ;
\draw [shift={(503,34)}, rotate = 89.59] [fill={rgb, 255:red, 0; green, 0; blue, 0 }  ][line width=0.08]  [draw opacity=0] (12,-3) -- (0,0) -- (12,3) -- cycle    ;
\draw  [line width=2.25]  (471.5,134.49) -- (502.75,103.25) -- (574.5,103.25) -- (574.5,179) -- (543.25,210.25) -- (471.5,210.25) -- cycle ; \draw  [line width=2.25]  (574.5,103.25) -- (543.25,134.49) -- (471.5,134.49) ; \draw  [line width=2.25]  (543.25,134.49) -- (543.25,210.25) ;

\draw (16,7) node [anchor=north west][inner sep=0.75pt]   [align=left] [font=\huge] {(a)};
\draw (372,6) node [anchor=north west][inner sep=0.75pt]   [align=left] [font=\huge] {(b)};
\draw (13,50) node [anchor=north west][inner sep=0.75pt]  [font=\Large]  {$\mathcal{X}_{0,1}$};
\draw (86,52) node [anchor=north west][inner sep=0.75pt]    {$0\leqslant x_{1} \leqslant 1$};
\draw (65,70) node [anchor=north west][inner sep=0.75pt]  [font=\footnotesize]  {$0$};
\draw (157,71) node [anchor=north west][inner sep=0.75pt]  [font=\footnotesize]  {$1$};
\draw (12,119) node [anchor=north west][inner sep=0.75pt]  [font=\Large]  {$\mathcal{X}_{0,2}$};
\draw (85,121) node [anchor=north west][inner sep=0.75pt]    {$0\leqslant x_{2} \leqslant 1$};
\draw (64,139) node [anchor=north west][inner sep=0.75pt]  [font=\footnotesize]  {$0$};
\draw (156,140) node [anchor=north west][inner sep=0.75pt]  [font=\footnotesize]  {$1$};
\draw (12,193) node [anchor=north west][inner sep=0.75pt]  [font=\Large]  {$\mathcal{X}_{0,3}$};
\draw (86,195) node [anchor=north west][inner sep=0.75pt]    {$0\leqslant x_{3} \leqslant 1$};
\draw (65,213) node [anchor=north west][inner sep=0.75pt]  [font=\footnotesize]  {$0$};
\draw (157,214) node [anchor=north west][inner sep=0.75pt]  [font=\footnotesize]  {$1$};
\draw (198,95) node [anchor=north west][inner sep=0.75pt]  [font=\huge]  {$ \begin{array}{l}
\xi =[ x_{1} ,x_{2} ,x_{3}]^{T}\\
\Xi _{0} =\mathcal{X}_{0,1} \times \mathcal{X}_{0,2} \times \mathcal{X}_{0,3}
\end{array}$};
\draw (405,235) node [anchor=north west][inner sep=0.75pt]    {$x_{3}$};
\draw (611,184) node [anchor=north west][inner sep=0.75pt]    {$x_{1}$};
\draw (477,24) node [anchor=north west][inner sep=0.75pt]    {$x_{2}$};

\end{tikzpicture}
    }
    \caption{(a) Distributed, and (b) centralized structures of the polytope}
    \vspace{-0.0em}
    \label{fig:plane-example}
\end{figure}
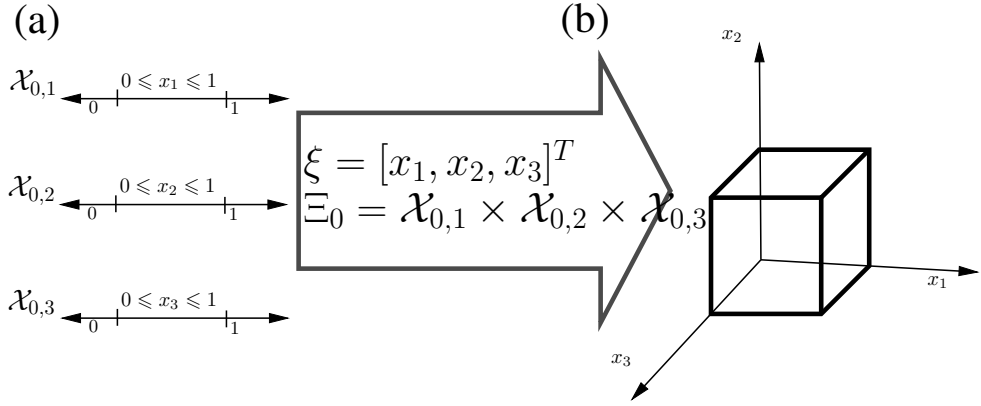

To observe the distributed nature of the parameters of the polytopes, consider a one-dimensional, 3 agent MAS case ($n=1$ and $N=3$).
The hyperplanes $F_i^k$ defining initial sets are simply inequalities, as shown in Fig.~\ref{fig:plane-example} (a).
The local hyperplanes are simply $F^1_i,F^2_i$, denoting the inequalities $x_1\geq 0$ and $x_i\leq 1$, respectively.
As a result, the centralized state $\xi$ lies in the cube of unit size as shown in Fig.~\ref{fig:plane-example} (b).

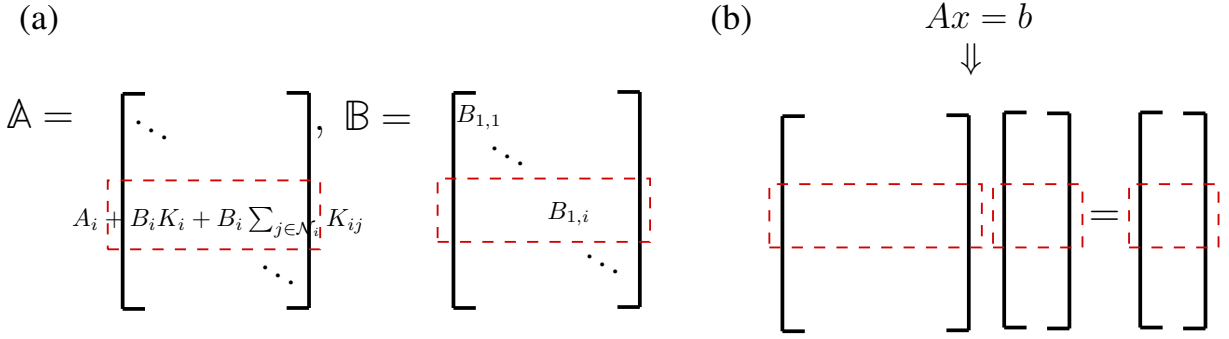
\begin{figure}[!t]
    \centering
    \resizebox{1\columnwidth}{!}
    {
    \tikzset{every picture/.style={line width=0.75pt}} 

\begin{tikzpicture}[x=0.75pt,y=0.75pt,yscale=-1,xscale=1]

\draw [line width=1.5]    (66,55.5) -- (66,152.5) -- (66,171) ;
\draw [line width=1.5]    (66,55.5) -- (78,55.5) ;
\draw [line width=1.5]    (66,171) -- (78,171) ;

\draw [line width=1.5]    (166,170.5) -- (166,73.5) -- (166,55) ;
\draw [line width=1.5]    (166,170.5) -- (154,170.5) ;
\draw [line width=1.5]    (166,55) -- (154,55) ;

\draw [line width=1.5]    (243.5,55) -- (243.5,152) -- (243.5,170.5) ;
\draw [line width=1.5]    (243.5,55) -- (255.5,55) ;
\draw [line width=1.5]    (243.5,170.5) -- (255.5,170.5) ;

\draw [line width=1.5]    (343.5,170) -- (343.5,73) -- (343.5,54.5) ;
\draw [line width=1.5]    (343.5,170) -- (331.5,170) ;
\draw [line width=1.5]    (343.5,54.5) -- (331.5,54.5) ;

\draw  [color={rgb, 255:red, 208; green, 2; blue, 2 }  ,draw opacity=1 ][dash pattern={on 4.5pt off 4.5pt}] (58,102) -- (172,102) -- (172,139) -- (58,139) -- cycle ;
\draw  [color={rgb, 255:red, 208; green, 2; blue, 2 }  ,draw opacity=1 ][dash pattern={on 4.5pt off 4.5pt}] (235,101) -- (349,101) -- (349,135) -- (235,135) -- cycle ;
\draw [line width=1.5]    (420,67.5) -- (420,164.5) -- (420,183) ;
\draw [line width=1.5]    (420,67.5) -- (432,67.5) ;
\draw [line width=1.5]    (420,183) -- (432,183) ;

\draw [line width=1.5]    (520,182.5) -- (520,85.5) -- (520,67) ;
\draw [line width=1.5]    (520,182.5) -- (508,182.5) ;
\draw [line width=1.5]    (520,67) -- (508,67) ;

\draw [line width=1.5]    (539,65.5) -- (539,162.5) -- (539,181) ;
\draw [line width=1.5]    (539,65.5) -- (551,65.5) ;
\draw [line width=1.5]    (539,181) -- (551,181) ;

\draw [line width=1.5]    (574,181.5) -- (574,84.5) -- (574,66) ;
\draw [line width=1.5]    (574,181.5) -- (562,181.5) ;
\draw [line width=1.5]    (574,66) -- (562,66) ;

\draw [line width=1.5]    (612,65.5) -- (612,162.5) -- (612,181) ;
\draw [line width=1.5]    (612,65.5) -- (624,65.5) ;
\draw [line width=1.5]    (612,181) -- (624,181) ;

\draw [line width=1.5]    (647,181.5) -- (647,84.5) -- (647,66) ;
\draw [line width=1.5]    (647,181.5) -- (635,181.5) ;
\draw [line width=1.5]    (647,66) -- (635,66) ;

\draw  [color={rgb, 255:red, 208; green, 2; blue, 2 }  ,draw opacity=1 ][dash pattern={on 4.5pt off 4.5pt}] (413,104) -- (527,104) -- (527,138) -- (413,138) -- cycle ;
\draw  [color={rgb, 255:red, 208; green, 2; blue, 2 }  ,draw opacity=1 ][dash pattern={on 4.5pt off 4.5pt}] (533,104) -- (581,104) -- (581,138) -- (533,138) -- cycle ;
\draw  [color={rgb, 255:red, 208; green, 2; blue, 2 }  ,draw opacity=1 ][dash pattern={on 4.5pt off 4.5pt}] (606,104) -- (654,104) -- (654,138) -- (606,138) -- cycle ;

\draw (9,6) node [anchor=north west][inner sep=0.75pt]  [font=\Large] [align=left] {(a)};
\draw (380,6) node [anchor=north west][inner sep=0.75pt]  [font=\Large] [align=left] {(b)};
\draw (2,56.5) node [anchor=north west][inner sep=0.75pt]  [font=\LARGE] [align=left] {$\displaystyle \mathbb{A} =$};
\draw (167,57.5) node [anchor=north west][inner sep=0.75pt]  [font=\LARGE] [align=left] {$\displaystyle ,\ \mathbb{B} =$};
\draw (243.5,60) node [anchor=north west][inner sep=0.75pt]  [font=\normalsize]  {$B_{1,1}$};
\draw (292.5,112) node [anchor=north west][inner sep=0.75pt]  [font=\normalsize]  {$B_{1,i}$};
\draw (260.5,73) node [anchor=north west][inner sep=0.75pt]  [font=\Large]  {$\ddots $};
\draw (311.5,131) node [anchor=north west][inner sep=0.75pt]  [font=\Large]  {$\ddots $};
\draw (69.5,59) node [anchor=north west][inner sep=0.75pt]  [font=\Large]  {$\ddots $};
\draw (137.5,137) node [anchor=north west][inner sep=0.75pt]  [font=\Large]  {$\ddots $};
\draw (37.5,115) node [anchor=north west][inner sep=0.75pt]  [font=\normalsize]  {$A_{i} +B_{i} K_{i} +B_{i}\sum _{j\in \mathcal{N}_{i}} K_{ij}$};
\draw (488,4) node [anchor=north west][inner sep=0.75pt]  [font=\Large]  {$ \begin{array}{l}
Ax=b\\
\ \ \ \ \Downarrow 
\end{array}$};
\draw (583,117) node [anchor=north west][inner sep=0.75pt]  [font=\LARGE] [align=left] {$\displaystyle =$};

\end{tikzpicture}
    }
    \caption{Distribution of system matrices information across the agents}
    \vspace{-0.0em}
    \label{fig:info-distr}
\end{figure}

Observe that due to the structure of $\Xi_0=\mathcal X_{0,1}\times \mathcal X_{0,2}\times \mathcal X_{0,3}$, each agent is aware of $n_1=2$ faces defining the polytope $\Xi_0$.
Each agent introduces $n_1$ number of planes to define $\Xi_0$, hence, $\Xi_0$ has $N n_1$ total faces.
The same is true for the polytope $\mathcal W$.
That is, \emph{the faces defining both sets, $\Xi_0$ and $\mathcal W$ are distributed across the graph $\mathcal G$} in the distributed reachability problem.
Similarly, the matrices $\mathbb A$ and $\mathbb B$ are also (row-wise) distributed across the graph $\mathcal G$ (see Fig.~\ref{fig:info-distr} (a)).
This distributed information structure affects all steps S1 through S3 previously described.
To this end, we need to carry out the steps S1 and S3 in a distributed manner, the `stacked state' for the $j^{\text{th}}$ hyperplane $\xi^*_j(\tau)$ and the corresponding `stacked co-state' $\lambda^*_j(\tau)$ evolution equations in (\ref{eq:support-point-evolve}), and (\ref{eq:costate-evolve}), respectively.
This can be achieved by converting the corresponding state evolution differential equations to linear equations using their Laplace transforms.
Note that for distributed linear dynamics (see Fig.~\ref{fig:info-distr} (a)), the corresponding Laplace transform of the dynamics is a distributed linear equation (d-LE) as (see Fig.~\ref{fig:info-distr} (b)) which can be written as:
\begin{align}
&\dot{\lambda}_j^*(\tau) = -\mathbb A^T{\lambda}_j^*(\tau) \xrightarrow{\mathcal L(\bullet)} s\Lambda^*_j(s) = -\mathbb A^T \Lambda^*_j(s) \label{eq:laplace} \\
& \Rightarrow (sI + \mathbb A^T)\Lambda^*_j(s) = \lambda^*_j(t_0) = c^*_j(t_0) \xrightarrow{\mathcal L^{-1}(\bullet)} \lambda^*_j(\tau) \label{eq:costate-d-leq}
\end{align}    
where $\Lambda^*_j(s)\triangleq \mathcal L(\lambda^*_j(t))$ is the Laplace transform of the co-state variable.
Therefore, the co-state equation is a d-LE in the Laplace domain, where each agent $i$ has information if its corresponding columns of the matrix based on local information set $\mathcal I_i$, as shown in Fig.~\ref{fig:info-distr} (b).
Similarly, for the Laplace transform of the state variable, $\mathcal{L}(\xi^*_j(t))\triangleq X^*_j(s), \mathcal{L}(W^*_j(t))\triangleq W^*_j(s)$, the stacked state evolves as:
\begin{align}\label{eq:state-d-leq}
(\ref{eq:support-point-evolve}) &\xrightarrow{\mathcal L(\bullet)}  (sI - \mathbb A)X^*_j(s) = \mathbb{B}W^*_j(s) + \xi^*_j(t_0)
\end{align}    
where $W^*_j(t)$ is given by the stacked optimal control law in (\ref{eq:optimal-control}).
Since each agent $i$ knows their own parameterizations for the local polytopes for $\mathcal X_{i,0},\mathcal W_i$, and their optimal control, $w^*j(\tau)$, the state and co-state propagation equations in (\ref{eq:costate-d-leq}), and (\ref{eq:costate-d-leq}), both admit a d-LE form.

Therefore, we first outline a distributed linear equation solving method adapted from \cite{mou2015distributed}, and next carry out the distribute optimal control step S2.
Consider a general d-LE as follows.
Solve $Ax=b$, where agent $i$ has access to certain rows of the problem, given by the tuple $(A_{[i,:]},b_i)$.
Additionally, the agents are connected according to the graph $\mathcal G$.
Then, the d-LE can be solved by each agent as:
\begin{equation}\label{eq:mou-LE-interation}
\hat{x}_i^{k+1} = \hat{x}_i^k - \proj{\mathrm{Ker\left(A_{[i,:]}\right)}}{\hat{x}_i^k - \frac{1}{\abs{\mathcal N_i}} \sum_{j\in\mathcal N_i} {\hat{x}_j^k}}
\end{equation} 
where $\proj{\mathbb{S}}{\pmb{s}}$ is the orthogonal projection of vector $\pmb{s}$ on to the set $\mathbb S$, and $\text{Ker}(\bullet)$ is the Kernel of the matrix $\bullet$ \cite{mou2015distributed}.
The d-LE solution method outlined above based on \cite{mou2015distributed} completes the distributed reachable set computation algorithm summarized in Algorithm \ref{algo:algo1}.
In the next section we analyze the convergence properties of Algorithm \ref{algo:algo1}.

\section{Convergence Discussion}\label{sec:convergence}
In order to study the convergence properties of the algorithm in the previous section for the general case of time-varying graphs, we first discuss time-invariant/static graph convergence properties.

\subsection{Convergence of Algorithm \ref{algo:algo1} for static graphs}
For the d-LE solution iterations in (\ref{eq:mou-LE-interation}), the following Lemma holds.
\begin{lemma}
If the graph $\mathcal G(t)$ is (repeatedly jointly) strongly connected, the iterations in (\ref{eq:mou-LE-interation}) converges to the solution to the equation $Ax=b$ exponentially fast.
\end{lemma}
\begin{proof}
See Theorem 1 from \cite{mou2015distributed}.
\end{proof}

Next, we carry out the step S2 to compute the optimal control in (\ref{eq:optimal-control}) across the graph $\mathcal G(t)$.
To compute $W^*_j(\tau)$, we can easily observe that we have polytopic bounds on the set $\mathcal W$ distributed among agents in the graph $\mathcal G$.
Note that the set $\mathcal W$ is a polytope, whose vertices are spread across agents in the MAS but can be computed offline (similar to Fig.~\ref{fig:info-distr} (b)).
Let the set $V\triangleq \{v_1,\cdots,v_K\}$ be the vertex set defining $\mathcal W$, and each agent be aware of a subset $V_i$ of the vertices (i.e., the vertex sets $V_i$`s partition $V$).
The problem of finding the optimal control $W^*_j(t)$ in (\ref{eq:optimal-control}) can be rewritten as:
\begin{equation}
W^*_j(t)=\argmax_{W_j(t)\in \text{convex hull}\{V\}} \langle \Phi_{-\mathbb{A}^T}(\tau,t_0) \lambda^*_j(t_0), W_j\rangle 
\end{equation}
However, the computation above is drastically simplified since Pontryagin's maximum principle guarantees that the maximum occurs on one of the vertices in $V$ \cite{varaiya2000reach}.
Therefore, each agent $i$ can simply exchange the maximum across their neighborhoods as:
\begin{equation}\label{eq:distr-vertex-optimalcontrol}
\begin{split}
\hat{W}^*_{i} &\gets \argmax_{v\in V_i}\{\langle \Phi_{-\mathbb{A}^T}(\tau,t_0) \lambda^*_j(t_0), v\rangle\} \\
\hat{W}^*_{i} &\gets \max_{j\in\mathcal N_i}{\{W^*_{i} , W^*_{j}\}}
\end{split}
\end{equation}
From (\ref{eq:distr-vertex-optimalcontrol}), agents in the MAS simply keep a track of the vertex in the set $V_i$ that maximizes the optimal control cost, and keep a local copy of the maximum of such vertices among its own neighborhood $\mathcal N_i$.
This simple procedure computes the maximum in at most $T$ steps, where $T$ is the graph diameter $\diam{\mathcal G}$.
Clearly, this requires the graph to be strongly connected, such that the local maximum is communicated through each neighborhood, and eventually the maximum is computed (in the worst case) along the longest path with the path length equaling the graph diameter.

This completes the fully distributed reachable set computation method, where we first proposed the centralized polytopic reachable set computation scheme and modified it to accommodate fully distributed MASs applications, \textit{for a time-invariant, undirected, strongly connected graph}.
For simplicity, we had assumed that the graph $\mathcal G(t)$ is strongly connected, and static at all times, i.e., $\mathcal G(t)=\mathcal G$.
We now present our main results for time-varying MAS graphs.

\subsection{Convergence of Algorithm \ref{algo:algo1} for Time-Varying Graphs}
We will consider the convergence of Algorithm \ref{algo:algo1} for a specific class of time-varying graphs.
To aid the convergence proof, we utilize the following properties from graph theory for sequences of graphs.
\begin{define}
Consider two graphs $\mathcal G_1,\mathcal G_2$ with the corresponding adjacency matrices $\adjm{\mathcal G_1}$ and $\adjm{\mathcal G_2}$, respectively.
Then their \emph{graph composition} is defined as the graph corresponding to the product of the two adjacency matrices as $\mathcal G_1 \circ \mathcal G_2 \triangleq \mathcal G(\adjm{\mathcal G_1}\adjm{\mathcal G_2})$.
\end{define}
\begin{define}\label{def:repeat-strong-conn}
Consider a (possibly infinite) time-varying graph sequence $\mathbb G\triangleq\{\mathcal G(t_1),\mathcal G(t_2),\cdots\}$.
Then the sequence $\mathbb G$ (see \cite{mou2015distributed}) is \emph{repeatedly jointly strongly connected}, if a over a finite interval $[\underline{t},\overline{t}]$ the graph composition $\mathcal G_{\circ [\underline{t},\overline{t}]}$ is strongly connected.
\end{define}

While strong connectivity of $\mathcal G$ requires the existence of a path between any two arbitrary agents in the graph $\mathcal G(t)$, a repeatedly-jointly connected graph sequence $\mathbb G$ requires the existence of a path between any two arbitrary agents over the time interval $[\underline{t}, \overline{t}]$.
Therefore, Definition \ref{def:repeat-strong-conn} is a generalization of the notion of strong connectivity, allowing for $\mathcal G(t)$ to be time-varying.
Based on the definitions above, we can consider the convergence of our proposed method \textit{for time-varying, undirected, repeatedly jointly strongly connected graphs}.
\begin{theorem}
Let the MAS under consideration evolve according to (\ref{eq:system-eqn}) where the information $\mathcal I_i$ is distributed across the graph $\mathcal G (t)$.
If the time-varying graph sequence $\mathbb G$ is repeatedly jointly strongly connected, then the reachable sets $\mathcal R_i(\tau;\mathcal X_{0,i},\mathcal W_i)$ can be computed efficiently. 
Specifically, the intermediate distributed computations can be performed exponentially fast (for steps S1 and S3) and the optimal control in S2 can be computed in finite time.
\end{theorem}
\begin{proof}
Similar to $\diam{\mathcal G}$ in (\ref{eq:distr-vertex-optimalcontrol}), the optimal control step for time-varying graphs now depends on the maximum diameter of the graph composition over the interval $[\underline{t},\overline{t}]$.
Hence, step S2 converges in at most $\sup_{t\in[\underline{t},\overline{t}]} \diam{{\mathcal G_{\circ [\underline{t},\overline{t}]}}}$ steps.
However, note that the slowest steps are still S1 and S3, both of which now require solving the distributed linear equations over the graph sequence $\mathbb G$.
From Theorem 1 in \cite{mou2015distributed}, the update rule for d-LE in (\ref{eq:mou-LE-interation}) converges exponentially fast to the solution of $Ax=b$ over repeatedly jointly strongly connected graph sequence $\mathbb G$.
Therefore, distributed solution to S2 converges to optimal control solution in finite time, while distributed solutions to S1 and S3 converge exponentially. 
\end{proof}

This completes the proof for exponential convergence of Algorithm \ref{algo:algo1} for time-varying graphs along with the conditions where the convergence results hold.
\begin{remark}
Note that the most computationally intensive steps in the distributed reachable set computation algorithm are \texttt{d-LP($\bullet)$}, and the Laplace transforms.
While \texttt{d-LP($\bullet$)} converges exponentially fast, the Laplace transforms for linear systems are as expensive as computing $\exp{\bullet}$.
Thus, both steps are numerically computable for the graphs described in the proofs above. 
\end{remark}

\begin{algorithm}[!t]
\setstretch{1.25}
\SetNoFillComment
\SetAlgoLined
\KwIn{$t_0$, $\tau, T$, local copies of $\mathcal I_i$ (for each agent $i\in \mathcal V$) from (\ref{eq:information})}
\Parameters{initial polytopes: $\{\lambda^*_j(t_0),c^*_j(t_0)\}_{j=1}^{n_1}, \{W^*_j\}_{j=1}^{n_2}$}
compute initial contact points $\{\xi^*_j(t_0)\}_{j=1}^{n_1}$\\
\While{$\tau\leq T$}
{
\For{$j\gets0$ \KwTo $n_1$}{
$\lambda_j^*(\tau) \gets$ co-state update using \texttt{dLP\_solver($\bullet$)} on (\ref{eq:costate-d-leq}) \\
}
\For{$j\in \mathcal V$}{
    solve for optimal control $\hat{W}^*_i$ using (\ref{eq:distr-vertex-optimalcontrol})
    }
\For{$j\gets 0$ \KwTo $n_1$}{
update $\xi^*_j(\tau)$ for $j^{\text{th}}$ hyperplane using $\lambda^*_j(\tau),W^*_j(\tau)$\\
$\xi^*_j(\tau) \gets$ \texttt{dLP\_solver($\bullet$)} on (\ref{eq:state-d-leq})
}
$\overline{\mathcal R_i}(\tau) \gets \{\lambda^*_i(\tau),\xi^*_i(\tau)\}$ \tcp*[l]{outer approx.}
$\underline{\mathcal R_i}(\tau) \gets \text{convex hull}\{\xi^*_i(\tau)\}$ \tcp*[l]{inner approx.}
}
{return}
\KwOut{$\{\overline{\mathcal R_i}(\tau;\mathcal X_{0,i},\mathcal W_i), \underline{\mathcal R_i}(\tau;\mathcal X_{0,i},\mathcal W_i)\}_{i\in \mathcal V}$}
\SetKwFunction{FdLP}{dLP\_solver}
\SetKwProg{Fn}{Function}{:}{}
\Fn{\FdLP{$(A_{[i,:]},b_i;i\in\mathcal V)$, $\mathcal G(t)$, $n_\varepsilon$}}{
$\hat{x}_i^0 \gets \pmb{0},\, \forall i\in \mathcal V$\tcp*[l]{arbitrary initialization for dLP solver}
\For{$k\gets0$ \KwTo $n_\varepsilon$}{
    Update $\hat{x}_i^{k+1}$ using (\ref{eq:mou-LE-interation})
    }
\KwRet{$x^*_i,\,\forall i\in\mathcal V$}
}
\caption{Distributed Reachable Set Computation}\label{algo:algo1}
\vspace{-0.0em}
\end{algorithm}


\section{Conclusions}\label{sec:conclusion}
In this work, we studied the problem of distributed reachable set computation for a multi agent system (MAS).
We studied the problem of computing polytopic approximations to the $\tau$-time reachable sets, and proposed a fully distributed version and computation scheme.
We employed techniques from distributed linear equation solving to carry out the corresponding distributed reachable set computations.
The proposed method was proven to obtain exponentially convergent reachable sets under communication graphs for the MAS that are repeatedly jointly strongly connected.

As our immediate future work, we will investigate convergence properties directed graphs for inter-agent communication in the MAS.
We will also look into more numerically efficient reachable set approximation methods (e.g., ellipsoidal reachability \cite{kurzhanski2000ellipsoidal}) in a fully distributed manner for time-varying MASs.

\bibliographystyle{plain}

\end{document}